\documentclass[reprint,amssymb,amsmath, aps, prl, cha, twocolumn]{revtex4-1}
\usepackage{graphicx}
\usepackage[caption = false]{subfig}
\usepackage{color}
\newcommand{\comment}[1]{}
\usepackage{epsfig}
\usepackage{ifpdf}
\usepackage{url}%
\usepackage{multirow}
\usepackage{float}
\usepackage{bm}
\usepackage[colorlinks=true,linkcolor=blue]{hyperref}%
\usepackage{cleveref}
\expandafter\ifx\csname package@font\endcsname\relax\else
\expandafter\expandafter
\expandafter\usepackage
\expandafter\expandafter
\expandafter{\csname package@font\endcsname}%

\usepackage{subcaption}
\usepackage{ragged2e}
\DeclareCaptionJustification{justified}{\justifying}
\fi
\begin{document}

\title{Auto-correlations of Microscopic Density Fluctuations for Yukawa Fluids  in  the Generalized Hydrodynamics Framework}
\author{Ankit Dhaka}
\email[]{dhaka.nitrkl@gmail.com}
\affiliation{Institute for Plasma Research, Bhat, Gandhinagar, Gujarat 382428, India}
\affiliation{Homi Bhabha National Institute, Training School Complex, Anushaktinagar, Mumbai 400094,India}

\author{PV Subhash}
\affiliation{ITER-India, Institute for Plasma Research, Bhat, Gandhinagar, Gujarat 382428, India}
\affiliation{Homi Bhabha National Institute, Training School Complex, Anushaktinagar, Mumbai 400094,India}

\author{P. Bandyopadhyay}
\affiliation{Institute for Plasma Research, Bhat, Gandhinagar, Gujarat 382428, India}
\affiliation{Homi Bhabha National Institute, Training School Complex, Anushaktinagar, Mumbai 400094,India}

\author{A. Sen}
\affiliation{Institute for Plasma Research, Bhat, Gandhinagar, Gujarat 382428, India}
\affiliation{Homi Bhabha National Institute, Training School Complex, Anushaktinagar, Mumbai 400094,India}

\date{\today}
\begin{abstract}
The present work develops a theoretical procedure for obtaining transport coefficients of Yukawa systems from density fluctuations. The dynamics of Yukawa systems are described in the framework of  the generalized hydrodynamic model that  incorporates strong coupling and visco-elastic memory effects.  A hydrodynamic matrix for such a system is exactly derived and  then used to obtain an analytic expression for  the density autocorrelation function (DAF)- a marker of the time dynamics of density fluctuations. 
The present approach is validated against a DAF  obtained from numerical data of  Molecular Dynamics (MD) simulations of a  dusty plasma system that is a practical example of  a Yukawa system. The MD results and analytic expressions derived from the model equations 
are then used to obtain various transport coefficients  and the latter are compared with values available in the literature from other models.  The influence of strong coupling and visco-elastic effects on the transport parameters are discussed. Finally, the utility of 
our calculations for obtaining reliable estimates of transport coefficients from experimentally determined DAF is pointed out.  

\end{abstract}

\maketitle

\section{\label{sec:intro} Introduction}

\paragraph A Yukawa system  generally consists of an ensemble of a large number of  charged particles embedded in an electrically neutral or quasi-neutral medium such that the bare charge of a particle is shielded by the medium particles. Yukawa systems have attracted a lot of research interest  because of their importance in many fields including space physics \cite{Coakley1992}, astrophysical systems \cite{Bliokh1995}, gas discharges \cite{Hariprasad2018,Couedel2009}, microelectronics, colloidal systems, the edge of thermonuclear fusion systems \cite{Winter1999,Tsytovich1998,Winter2000}, condensed matter physics (specially for understanding the phase transitions \cite{Chu1994,Vaulina2006,Hariprasad2020} in 2D and 3D systems), etc. Such systems are also extensively studied in laboratories to investigate various fundamental physics problems associated with many body systems \cite{Thomas1994,Ivlev2003,Schweigert1996}. As a large amount of information is already available in the literature regarding  the domain of existence and the fundamental importance of Yukawa systems further details about them are omitted here. Good overviews of their basic properties,  applications, and methods of experimental and theoretical studies of Yukawa systems can be found in several  books and review papers [\onlinecite{Shukla1992,Shukla2009,Shukla_2002,Melzer2019}]. Complex plasmas or dusty plasmas are a particular class of Yukawa systems where nano-meter to micro-meter sized charged particles (called dust) are suspended in a partially ionized plasma.  Many past studies have investigated transport processes, crystallization, phase transitions and collective modes in  Yukawa systems  using various approaches such as Molecular Dynamics (MD) simulations \cite{Salin2003,Vaulina2008}, a Generalized Hydrodynamics (GH) model \cite{Kaw1998}, a Quasi-Localized Charge Approximation (QLCA) \citep{Golden2000} and Kinetic Theory \cite{Murillo2000} etc. 

In a Yukawa system the inter-particle shielded potential between the embedded grains  is taken to be of the form:
\begin{equation}
\phi(r) = \frac{Q^2}{4\pi\epsilon_0}\frac{\exp(-r/\lambda_D)}{r}
\label{eq:yukawa}
\end{equation}
where $r$ is the separation between two particles having charge $Q$, $\epsilon_0$ is the permittivity of free space and $\lambda_D$ is the screening length arising from the background plasma. Yukawa systems can be characterized by two dimensionless parameters, namely,  the Coulomb coupling strength defined as $\Gamma = Q^2/(4\pi\epsilon_0a_{ws}k_BT_D)$  and the screening strength defined as $\kappa = a_{ws}/\lambda_D$ where $a_{ws}$ is the average inter-particle distance, $T_D$ is the temperature and $k_B$ is Boltzmann's constant. The Coulomb coupling parameter and screening strength can be adjusted to achieve longer or shorter correlations among the particles, that can characterize  the phase state of the system as being a fluid or a solid~ \cite{Hamaguchi1997}. 

\paragraph{Density Fluctuations:} The time evolution of small density fluctuations of fluids around the equilibrium values can be used to understand the transport process at a fundamental level. This was famously noted way back by Landau-Placzek \cite{mlandau1934}, who had observed that the variation of density fluctuations in time can be described by linear hydrodynamic equations of irreversible thermodynamics. A similar  statement by R.Kubo that \lq\lq the linear response of a given system to an external perturbation is expressed in terms of fluctuation properties of the system in thermal equilibrium"  is also noteworthy \cite{mKubo_1966}.  One way to understand the time evolution of fluctuations is to write down conservation laws such as conservation of density, momentum, and energy in the hydrodynamic limit with quantities having small fluctuations around their equilibrium values. After linearising the equations one can further simplify them using thermodynamic relations to reach a set of coupled equations. This set of equations relates fluctuations of density, momentum, and energy to their equilibrium values. The system of equations when written in matrix form has a coefficient matrix, which is normally called the hydrodynamic matrix \cite{Hansen2013}. Following some reasonable assumptions, these equations can be solved for variation of density fluctuations in time in terms of various equilibrium values. To understand the time dynamics of density fluctuations, a time auto-correlation function of density fluctuation can be constructed. This is found to yield much important information on transport processes in fluids. Such a calculation for the case of 'simple fluids' can be found in Ref [\onlinecite{Hansen2013}]. This observation has been implemented in the light scattering studies from ideal mono-atomic liquids by \citet{mMOUNTAIN1966}, to construct the generalized structure factor and other dynamical quantities. Following the work of \citet{mMOUNTAIN1966}, the same approach has been used to study thermodynamic density fluctuations for a dense charged fluid (a strongly coupled one component plasma (OCP)) by \citet{Vieillefosse1975}. They added a local electric field term in the momentum equation to incorporate the effects of charged particles. This procedure to understand transport parameters is more accurate as one starts from an unambiguous quantity, the density fluctuations, and is valid for complicated situations like materials with non-pairwise potentials such as warm dense matter etc.~\cite{Cheng2020}. 

Recent advances in Molecular Dynamic (MD) simulations give another dimension to this method. Using such simulations, we can numerically calculate the density fluctuations and hence the Density Autocorrelation Function (DAF). This numerically constructed DAF can then be compared with theoretical DAF to obtain important transport parameters and acoustic speeds. Recently \citet{Cheng2020} used this combination to successfully calculate transport parameters of simple fluids as well as for warm dense matter.

The situation in Yukawa fluids is more interesting as there exists a third possibility to obtain DAF through experiments. Dusty plasmas, a particular class of Yukawa fluids, are extensively studied in the laboratory  and their dynamics are captured in the form of particle trajectories using high speed camera systems \cite{mMelzer1996,mGoree2004,mThomas2005,Jaiswal2016,Hariprasad2018}. To exploit this aspect we need to have an accurate expression for DAF derived from a proper hydrodynamic matrix for Yukawa systems. It can be noted from a comparison between Ref [\onlinecite{Cheng2020}] and Ref [\onlinecite{Vieillefosse1975}] that the DAFs, hence the transport parameters, are different for simple fluids and OCP because of the additional term in the momentum equation. In Yukawa fluids, strong coupling effects as well as visco-elastic effects (sometimes called memory effects) need to be incorporated in the fluid equations. The Generalized Hydrodynamics model is one such model that incorporates both these features \cite{Berkovsky1992}.  This model has been applied  to a dusty plasma system by  \citet{Kaw1998} for studying low-frequency dust acoustic modes.   

In the present work, we derive an appropriate hydrodynamic matrix and density autocorrelation function in the framework of GH model and validate our results with MD simulations. Various physical insights obtained from the analytical DAF,  especially the presence of strong coupling effects and visco-elastic effects on DAF, various transport parameters, and sound speed will be elucidated with the help of MD simulations.

\section{\label{sec:model}Hydrodynamic Matrix and DAF from Generalized Hydrodynamic Model}
\subsection{Hydrodynamic Matrix}
Assuming the hydrodynamic regime, the conservation laws for number density $\rho (\bm{r},t)$ and energy density $e(\bm{r},t)$ can be written as
\begin{equation}
m\frac{\partial}{\partial t}\rho (\bm{r},t) + \bm{\nabla}\cdot \bm{p}(\bm{r},t) = 0
\label{eq:continuity}
\end{equation}
\begin{equation}
\frac{\partial}{\partial t}e(\bm{r},t) + \bm{\nabla}\cdot \bm{J}^e(\bm{r},t) = 0
\label{eq:econtinuity}
\end{equation}

where   $\bm{J}^e(\bm{r},t)$ is the energy current density and $\bm{p}(\bm{r},t)$ is the momentum current density. Now, assuming the local deviation in number density $\delta \rho(\bm{r},t)$ to be small, Eq.~(\ref{eq:continuity}) can be linearised as 
$$
\bm{p}(\bm{r},t) = m[\rho +\delta \rho(\bm{r},t)]\bm{u}(\bm{r},t)\approx m\rho\bm{u}(\bm{r},t) \equiv m\bm{j}(\bm{r},t)
$$

with $m$ as mass, $\bm{u}(r,t)$ as velocity and $\bm{j}(\bm{r},t)$ as the local density current. Using the above approximation, the continuity Eq.~(\ref{eq:continuity}) can be rewritten in the form
\begin{equation}
\frac{\partial}{\partial t}\delta \rho (\bm{r},t) + \bm{\nabla}\cdot \bm{j}(\bm{r},t) = 0
\label{eq:continuity2}
\end{equation}

Considering the heat continuity Eq.~(\ref{eq:econtinuity}), the heat current is defined as 
$$\bm{J}^e(\bm{r},t) = (e+P)\bm{u}(\bm{r},t) - \lambda \nabla T(\bm{r},t)$$ 
where $e=U/V$ is the equilibrium energy density, $\lambda$ is thermal conductivity and $P$ is the overall pressure. Using the expression for $J^e$, the \textit{energy equation} (\ref{eq:econtinuity}) can be rewritten as,
\begin{equation}
\frac{\partial}{\partial t} \delta \underbrace{ \left( e(\bm{r},t) - \frac{e+P}{\rho}\rho(\bm{r},t) \right)}_{\text{Density of Heat Energy : }q(\bm{r},t)} - \lambda\nabla^2\delta T(\bm{r},t) = 0
\label{eq:heat_energy_diffusion} 
\end{equation}
$\delta q(\bm{r},t)$ is related to $\delta \rho (\bm{r},t)$ and $\delta T(\bm{r},t)$ as,
\begin{eqnarray}
\delta q(\bm{r},t) = T\delta s(\bm{r},t) = &&\frac{T}{V}\frac{\partial S}{\partial \rho}\delta \rho (\bm{r},t)+ \frac{T}{V}\frac{\partial S}{\partial T}\delta T(\bm{r},t) \nonumber \\= &&-\frac{T\beta_v}{\rho}\delta \rho (\bm{r},t)+\rho c_v \delta T(\bm{r},t)
\label{eq:heat_energy} 
\end{eqnarray}
where $\beta_v = \left( \frac{\partial P}{\partial T}\right)_\rho = -\rho \left( \frac{\partial (S/V)}{\partial \rho}\right)_T $ is the thermal pressure coefficient. Using Eqs. (\ref{eq:continuity2}) \& (\ref{eq:heat_energy}), the energy continuity equation (\ref{eq:heat_energy_diffusion}) can be rewritten as
\begin{equation}
\label{eq:en_final}
\left( \frac{\partial}{\partial t} -\frac{\lambda}{\rho c_v}\nabla^2 \right)\delta T(\bm{r},t) + \frac{T\beta_v}{\rho^2 c_v}\bm{\nabla}\cdot \bm{j}(\bm{r},t) = 0
\end{equation}
\par The use of equilibrium  thermodynamic relations in Eq.~(\ref{eq:heat_energy}) is justified on the same grounds as the use of irreversible hydrodynamic equations to describe the time evolution of reversible microscopic fluctuations \cite{mlandau1934,mKubo_1966,Hansen2013}. In other words, the irreversibility  is at the macroscopic scale of the transport processes but there exists reversibility at the local microscopic scale of the fluctuations. 

The momentum conservation equation from Generalized Hydrodynamic model \cite{Kaw1998} can be written as

\begin{eqnarray}
&&\left(1+\tau_m \frac{\partial }{\partial t} \right)\left[ \frac{\partial }{\partial t}\bm{j}(\bm{r},t) + \frac{1}{m}\nabla P(\bm{r},t) + \frac{Q\rho}{m}\nabla \phi \right]-\nonumber \\  &&\frac{\eta}{\rho m } \nabla^2 \bm{j}(\bm{r},t) -\frac{\eta/3 +\zeta}{\rho m}\nabla \bm{\nabla} \cdot \bm{j}(\bm{r},t)=0
\label{eq:navstock}
\end{eqnarray}
with  $\eta$ as shear viscosity and $\zeta$ as the bulk viscosity.
Fluctuations in $P(\bm{r},t)$ to first order in $\delta \rho(\bm{r},t)$ and $\delta T(\bm{r},t)$ are related as
\begin{equation}
\delta P (\bm{r},t) = \frac{1}{\chi_T \rho}\delta \rho (\bm{r},t)+\beta_v\delta T(\bm{r},t) 
\label{eq:p_fluc}
\end{equation}

where $\chi_T$ is isothermal compressiblity.  Using Eq.~(\ref{eq:p_fluc}), the momentum, energy and continuity equation, for a Yukawa system,  can be rewritten as
\begin{eqnarray}
&&\left(1+\tau_m \frac{\partial }{\partial t} \right) \left[ \frac{\delta \rho(\bm{r},t)}{m\chi_T \rho} +\frac{\beta_v}{m}\delta T(\bm{r},t) +\frac{Q\rho}{m}\nabla \phi \right  ]+\nonumber \\ 
&&\left\{ \left(1+\tau_m \frac{\partial }{\partial t} \right) \frac{\partial }{\partial t} -\frac{\eta}{\rho m } \nabla^2 -  \frac{\eta/3 +\zeta}{\rho m}\nabla \bm{\nabla} \cdot  \right \}\bm{j}(\bm{r},t)=0 \nonumber \\
\label{eq:gh_momt}
\end{eqnarray}
\begin{equation}
\left( \frac{\partial}{\partial t} - \frac{\lambda}{\rho c_v}\nabla^2 \right)\delta T(\bm{r},t) + \frac{T\beta_v}{\rho^2 c_v}\bm{\nabla}\cdot \bm{j}(\bm{r},t) = 0
\label{eq:energy_cont}
\end{equation}
\begin{equation}
\frac{\partial}{\partial t}\delta \rho (\bm{r},t) + \bm{\nabla}\cdot \bm{j}(\bm{r},t) = 0
\label{eq:den_fluc_onti}
\end{equation}
\par The relation between density $\rho$ and $\phi$ can be established by using a  modified Helmholtz \cite{rWinkelmann2021} like  equation that is a static version of Eq.~(3) of  reference \cite{rYukawa}. 
\begin{equation}
(\nabla ^2 -\lambda_D^{-2})\phi = 4\pi Q \delta \rho
\label{eq_modihelmohltz}
\end{equation}
\par This equation relates the potential induced due to variation in the charge density ($Q\delta \rho$) for the system interacting via Yukawa interaction.
Now, the GH momentum equation (Eq. \ref{eq:gh_momt}) along with particle and energy conservation laws (Eq.~\ref{eq:energy_cont} \& Eq.~\ref{eq:den_fluc_onti}) can be transformed using a double transform with respect to space (Fourier) and time (Laplace) to obtain a relation of density $\tilde{\rho}(\bm{k},s)$, particle current density $\tilde{\bm{j}}(\bm{k},s)$ and local temperature $\tilde{T}(\bm{k},s)$ with their corresponding Fourier components, $\rho_k$, $T_k$ and $\bm{j}_k$, at $t=$0. The Laplace transform of function $f(t)$ has the form $ \mathcal{L}\{f(t)\} = \int \exp(\iota s t)f(t)dt$. Assuming $\bm{k}$ to be  in the $z$ direction (without losing generality) and neglecting electromagnetic effects, the longitudinal part of Eqs. (\ref{eq:gh_momt}), (\ref{eq:energy_cont}) and (\ref{eq:den_fluc_onti}) can be written in $(k,s)$ space as follows 

\begin{eqnarray}
-\iota s\tilde{\rho}_{k}(s)+ \iota k \cdot \tilde{j}_{kz}(s)=\rho_{k}(0)\nonumber \\ 
(-\iota s + ak^2)\tilde{T}_{k}(s) + \frac{T\beta_v}{\rho^2 c_v} \iota k \tilde{j}_{kz}(s)  = T_{\bm{k}}(0) \nonumber \\
\frac{\iota k\tilde{\rho}_{k}(s)}{1-\iota s \tau_m} \left[ \frac{1}{m\chi_T \rho} +\frac{\omega_p^2}{k^2+\lambda_D^{-2}}\right] +\frac{\iota k \beta_v(1-a\tau_mk^3)}{m (1-\iota s \tau_m)} \tilde{T}_{\bm{k}}(s)+ \nonumber \\
\frac{\tilde{j}_{kz}(s)}{1-\iota s \tau_m} \left( \left[ \frac{1}{m\chi_T \rho} + \frac{\beta_v^2}{\rho^2 m c_v} \right]k^2\tau_m +bk^2 -(\iota s+\tau_ms^2 )  \right)\nonumber \\ 
= j_{kz}(0) + \frac{\tau_m \dot{j_{kz}}(0)}{1-\iota s \tau_m} \nonumber
\end{eqnarray}

where $b=\frac{4\eta/3 +\zeta}{\rho m}$ and $a = \frac{\lambda}{\rho c_v}$ the thermal diffusivity.  The above equations can be written in matrix form as follows.
\begin{widetext} 
\begin{equation}
\underbrace{
\begin{bmatrix}
-\iota s &  0 & \iota k \\ 
0 &  -\iota s + ak^2& \frac{T\beta_v \iota k}{\rho^2 c_v} \\
\frac{\iota \bm{k}\tilde{\rho}_{\bm{k}}(s)}{1-\iota s \tau_m}\left[ \frac{1}{m\chi_T \rho} +\frac{\omega_p^2}{k^2+\lambda_D^{-2}}\right]  & \frac{\iota \bm{k} \beta_v(1-a\tau_mk^3)}{m (1-\iota s \tau_m)} 
& 
\begin{matrix} \frac{1}{(1-\iota s \tau_m)}\left ( \left [ \frac{1}{m\chi_T \rho} + \frac{\beta_v^2}{\rho^2 m c_v} 
\right ] k^2\tau_m\right) + \\ \hfill{ } \frac{bk^2 -(\iota s+\tau_ms^2) }{(1-\iota s \tau_m)} 
\end{matrix}
\end{bmatrix}
}_{\text{Hydrodynamic Matrix : }H_L(s,k)}
\begin{bmatrix}
\tilde{\rho}_k(s) \\
\tilde{T}_k(s) \\
\tilde{j}^z_k(s) \\
\end{bmatrix}
=
\begin{bmatrix}
\rho (0) \\
T (0) \\
{j}^z_k (0) + \frac{\tau_m \dot{j_{kz}}(0)}{1-\iota s \tau_m}\\
\end{bmatrix}
\label{eq:ghhydromatrix}
\end{equation}
\end{widetext}
The coefficient matrix in Eq.~(\ref{eq:ghhydromatrix}) is called the Hydrodynamics matrix $H_L(s,k)$. 

\subsection{Density Autocorrelation function (DAF)}
The dispersion relation for the longitudinal collective modes is determined by the poles of the inverse of $H_L(s,k)$ i.e. the roots of Eq.~(\ref{eq:detHLeq}). 
\begin{equation}
\det{H_L(s,k)}=0
\label{eq:detHLeq}
\end{equation}

Assuming $-\iota s = z$, $\det{H_L(s,k)}$ can be written as follows
\begin{eqnarray}
&&\det{H_L(s,k)}=z^4\tau_m + z^3(1+a\tau_mk^2)+ \nonumber \\ 
&&z^2k^2(a+b+\tau_m K_T) + zk^2\left[ ak^2\left(b+\frac{K_T\tau_m}{\gamma} \right) +\frac{\omega_p^2}{k^2+\lambda_D^{-2}}\right] \nonumber \\
&&+ak^4\left(\frac{K_T}{\gamma}+\frac{\omega_p^2}{k^2+\lambda_D^{-2}} \right )
\label{eq:detHL}
\end{eqnarray}
where definitions of $K_T$ as $K_T=\frac{\gamma}{\rho m \chi_T}$ and thermodynamic relation $c_p = c_v +\frac{T\chi_T \beta_v^2}{\rho}$ have been used.
 
The approximate roots of Eq.~(\ref{eq:detHLeq}) of the order $k^2$ can be determined using power series method as shown in Appendix \ref{sec:appendix}, as follows.

\begin{eqnarray*}
z_1 &&= \underbrace{-a \left ( 1 + \frac{K_T(1/\gamma -1)}{K_T + \frac{\omega_p^2}{k^2+\lambda_D^{-2}}}\right )}_{A}k^2 \\
z_{2\pm}&&=\pm\iota\underbrace{\sqrt{K_T + \frac{\omega_p^2}{k^2+\lambda_D^{-2}}}}_{c_s}k - \\
&&\underbrace{\frac{1}{2}\left(b+a\frac{K_T(1-1/\gamma)}{K_T + \frac{\omega_p^2}{k^2+\lambda_D^{-2}}} -\frac{\omega_p^2\tau_m}{k^2+\lambda_D^{-2}} \right)}_{\Gamma_s}k^2\\
z_4 &&= -\frac{1}{\tau_m} + \left[b-\frac{\omega_p^2 \tau_m}{k^2+\lambda_D^{-2}}\right]k^2
\end{eqnarray*}
As the fluctuations in temperature and density are instantaneously  uncorrelated \cite{Hansen2013} ($\left\langle T_k \rho_k\right\rangle=0$) and $\bm{k}$ can be chosen to ensure $j_k^z=0$. Considering these simplifications, the Eq.~(\ref{eq:ghhydromatrix}) can be solved for $\tilde{\rho}_{\bm{k}}(s)$
\begin{eqnarray*}
\frac{\tilde{\rho}_{k}(s)}{\rho_{k}}=&&\frac{z^3\tau_m +z^2(1+a\tau_mk^2+zk^2(a+b+\tau_m K_T))}{\tau_m (z-z_1)(z-z_{2+})(z-z_{2-})(z-z_4)}\\
&&+\frac{(\gamma-1)K_Tk^2/\gamma}{\tau_m (z-z_1)(z-z_{2+})(z-z_{2-})(z-z_4)}\\
\end{eqnarray*}

Using the roots of $\det H_L(s,k)=0$, we can solve for $\tilde{\rho}_k$ by finding partial fraction coefficients corresponding to each root. As we show later,  the coefficient corresponding to the fourth root in the density autocorrelation will be zero so the same is excluded from here onwards.

\begin{eqnarray*}
\frac{\tilde{\rho}_{k}(s)}{\rho_{k}} = &&\left( \frac{K_T(\gamma-1)/\gamma}{K_T+\frac{\omega_p^2}{k^2+\lambda_D^{-2}}} \right)\frac{1}{z-z_1}+\\
&&\frac{1}{2}\left(1-\frac{K_T(\gamma-1)/\gamma}{K_T+\frac{\omega_p^2}{k^2+\lambda_D^{-2}}} \right)\left(\frac{1}{z-z_{2+}} + \frac{1}{z-z_{2-}}\right)  \\
&&
\end{eqnarray*}

Now, above equation can be written as following using an inverse transform.  
\begin{eqnarray}
{\rho}_{k}(t)/\rho_{k}(0) = && \left(\frac{K_T(\gamma-1)/\gamma}{K_T+\frac{\omega_p^2}{k^2+\lambda_D^{-2}}} \right) \exp\left (-Ak^2t\right ) +\nonumber \\
&&\left(1-\frac{K_T(\gamma-1)/\gamma}{K_T+\frac{\omega_p^2}{k^2+\lambda_D^{-2}}} \right)\exp(-\Gamma_sk^2t)\cos (c_skt)\nonumber \\
\end{eqnarray}
Multiplying both sides by $\rho_{-k}(0)$ and taking thermal averages, the Density Autocorrelation Function (DAF) can be obtained to a second order approximation in $k$ as below :
\begin{eqnarray}
\left \langle \rho_k (t) \rho_{-k}(0) \right\rangle &&=\left(\frac{K_T(\gamma-1)/\gamma}{K_T+\frac{\omega_p^2}{k^2+\lambda_D^{-2}}} \right) \exp\left (-Ak^2t\right )+\nonumber \\
&& \left(1-\frac{K_T(\gamma-1)/\gamma}{K_T+\frac{\omega_p^2}{k^2+\lambda_D^{-2}}} \right)\exp(-\Gamma_sk^2t)\cos (c_skt)\nonumber \\
\label{eq_ghdenflu}
\end{eqnarray}

The attenuation constant $\Gamma_s$, coefficient $A$ and acoustic speed $c_s$ are given by
\begin{eqnarray}
\Gamma_s = &&\frac{1}{2}\left(b-\frac{aK_T(1/\gamma-1)}{K_T + \frac{\omega_p^2}{k^2+\lambda_D^{-2}}} -\frac{\omega_p^2\tau_m}{k^2+\lambda_D^{-2}} \right)\\ 
\label{eq:Gamma_s}
A = &&a\left(1+\frac{K_T(1-\gamma)/\gamma}{K_T+\frac{\omega_p^2}{k^2+\lambda_D^{-2}}} \right) \\
c_s = &&\sqrt{K_T + \frac{\omega_p^2}{k^2+\lambda_D^{-2}}}
\label{eq:sound_speed}
\end{eqnarray}

It can be noted immediately that putting t=0 in Eq.~(\ref{eq_ghdenflu}) reduces DAF to unity as expected, which in turn needs the coefficient of the fourth root to be zero.

The Eq.~(\ref{eq_ghdenflu}) contains two terms, the first one is a diffusive term driven by thermal diffusion and the second term is a damping cosine. The frequency of the cosine is described by sound speed, and the decay rate is determined through attenuation constant $\Gamma_s$, hence called sound attenuation constant. 

In the asymptotic limit of $\lambda_D \rightarrow \infty$ and $\tau_m \rightarrow 0$, Eq.~(\ref{eq_ghdenflu}) will recover the density autocorrelation function of classical one-component plasma (without memory effects) as shown by \citet{Vieillefosse1975}.
\subsubsection*{Extension to other fluids}

An important aspect of the present study is that it is possible, in principle, to extend the derivation to other fluid systems  that have interaction potentials different from the present Yukawa potential. The essential steps for such an extension are to calculate the force term from the given potential for use in the generalized momentum equation (Eq. \ref{eq:navstock}) and furthermore provide a closure relation to replace the modified Helmholtz equation (\ref{eq_modihelmohltz}).  For example, if we take the interaction potential to be a pure Coulomb one, namely, 
$$\phi = \frac{Q}{4\pi \epsilon_0 r}$$
 then the force term will  $F= -Q\nabla \phi = Q^2 / (4\pi \epsilon_0 r^2)$ and the closure relation is the Poisson equation,
$$ \nabla ^2 \phi = 4\pi Q\delta \rho(\bm{r}) $$
Another example would be an interaction potential of a Yukawa fluid that also takes account of effects arising from an overlap of the Debye spheres of the interacting particles. In such a case the interaction potential is given by \citet{mResendes_1998},
\begin{equation}
\phi = Q e^{-r / \lambda_D}\left(\frac{1}{r} -\frac{1}{2} \right)
\label{LJ}
\end{equation}
The force term to be used in the momentum equation is then 
$$F= -Q\nabla \phi = Q^2e^{-r /\lambda_D} \left( \frac{1}{r^2} +\frac{1}{r \lambda_D} - \frac{1}{2 \lambda_D} \right) $$
The potential in Eq.~(\ref{LJ}) has features similar to that of a Lennard-Jones potential in that the force is strongly repulsive at short distances and weakly attractive at larger distances.  The corresponding closure relation is given 
by the equation,
\begin{equation}
\nabla^2 \phi   -\left ( \lambda_D^{-2} -\lambda_D^{-1}\frac{2}{r-2} \right)\phi = 4\pi Q\delta\rho(\bm{r}) 
\label{clos2}
\end{equation}
Note that Eq.~(\ref{LJ}) is the Green's function solution of Eq.~(\ref{clos2}).

\section{Validation with MD Simulations}
\subsection{Calculation of DAF through MD}

The DAF  described in Eq.~(\ref{eq_ghdenflu}) can be independently calculated through the first principle method using molecular dynamics (MD) simulations. The MD simulations numerically solve the coupled equations of motion of particles for a given inter-atomic force field. As the solver progresses in time, the dynamical evolution of the system is recorded by storing the position and velocities of particles, also known as trajectories. This in turn produces a full 6N+1 dimensional phase space of the system with N being the number of particles.    The physical observables can now be calculated from this data with various statistical tools.

In the present study we have performed MD simulations of N=131072 point-like particles using a well benchmarked and well-established MD code LAMMPS \cite{Plimpton1995} using the Yukawa potential as in Eq.~(\ref{eq:yukawa}) for the inter-particle potential. The particle  number N has been chosen considering value of $k_{min}a_{ws} = \sqrt{4\pi/N}$ as per the $O(k^2)$ assumption in theoretical model. A periodic boundary condition is implemented to minimize the finite size effects. The system is initially equilibrated using a thermostating procedure \cite{Hoover1996} followed by a NVE production run of 60000 $\omega_{pd}t$ time steps for storing particle trajectories. The $\omega_{pd}$ here is the dust plasma frequency given by $\omega_{pd}=\sqrt{nQ^2/m\epsilon_02a_{ws}} $. The lengths and times are normalized with $a_{ws}$ and $2\pi\omega_{pd}^{-1}$ respectively. All other quantities are also normalized using the normalization scheme employed for time and space. As the potential given in Eq.~(\ref{eq:yukawa}) falls as $r$ increases, a potential truncation radius is used to speed up the computation which is chosen as per a benchmarked criteria explained by \citet{Liu2005}.

In order to calculate DAF from particle trajectories, the microscopic particle density in the reciprocal space for N point particles with positions $\bm{r}_i(t)$ is defined as
\begin{equation}
{\rho}_{\bm{k}}(t) = \frac{1}{V}\int_V \sum_{i=0}^N\delta (\bm{r}-\bm{r}_i(t))\exp(\iota\bm{k}\cdot\bm{r})d\bm{r}
\label{eq:den}
\end{equation}

 The reciprocal space vector $\bm{k}$ is related to system dimensions $(L_x,L_y,L_z)$ as $\bm{k} = \{ 2\pi n_x/L_x,2\pi n_y/L_y,2\pi n_z/L_z \}$.  Now, the microscopic particle density ${\rho}_{\bm{k}}(t)$ is transformed with a Fast Fourier Transform (FFT) in time and Wiener-Khintchine ~\cite{mKhintchine1934} theorem is used to calculate density autocorrelation in time using following equation.
\begin{equation}
\left \langle \rho_k (t) \rho_{-k}(0) \right\rangle= \mathcal{F}^{-1} \{ \tilde{ \rho}_k(\omega)^{\dagger}\tilde{\rho}_k(\omega)\}
\end{equation}

\begin{figure}
\includegraphics[width=\linewidth]{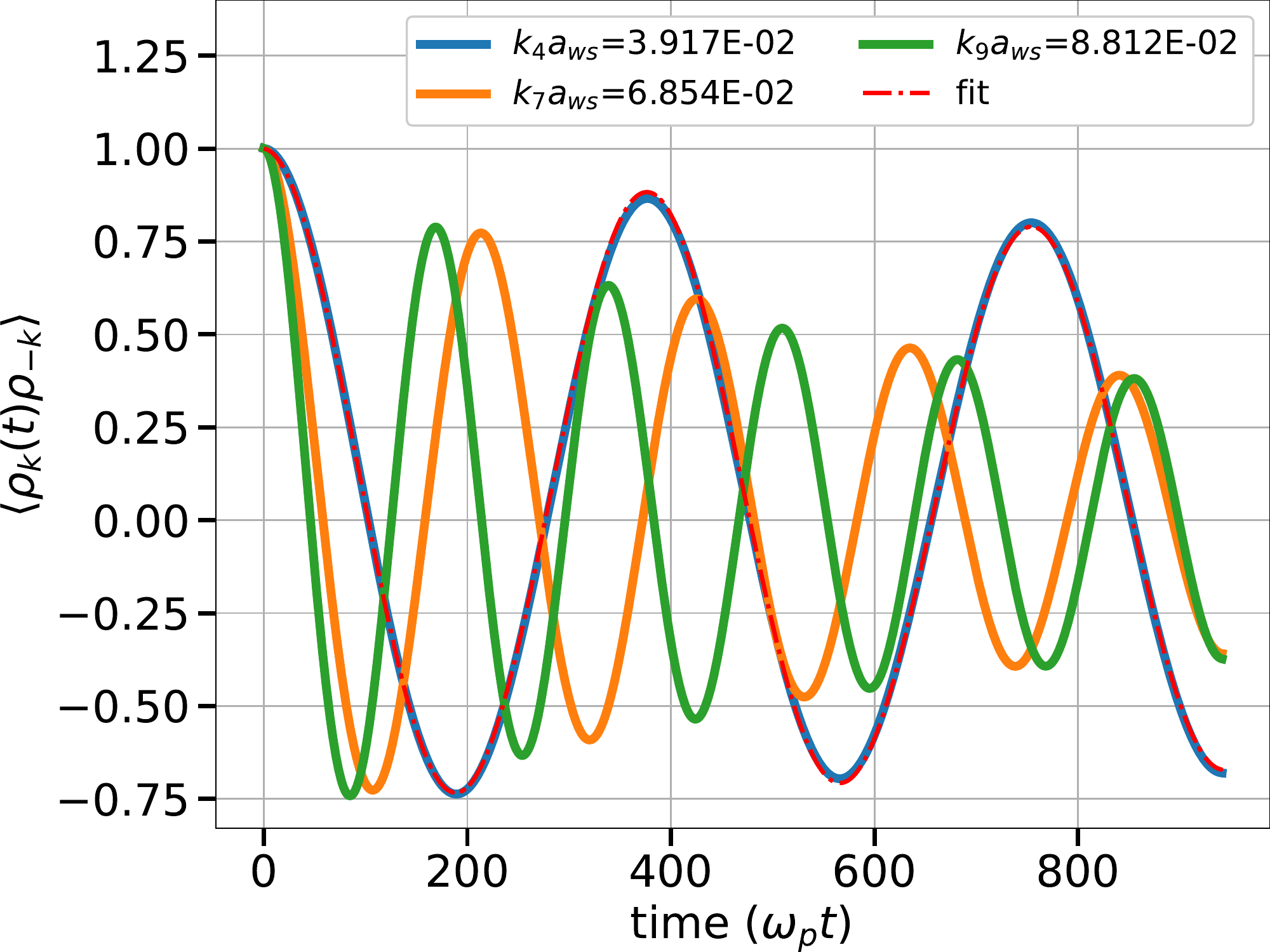}%
\caption{DAF curves generated using MD simulations (solid lines) and curve obtained by fitting MD data with Eq.~(\ref{eq_ghdenflu}) (broken lines) for $\Gamma$=60 and $\kappa$=2}.
\label{fig:ckt_fit}
\end{figure}

The DAF calculated from the simulation for $\Gamma$=60 and $\kappa$= 2.0 is shown in Fig.~\ref{fig:ckt_fit} (solid lines) for 3 different modes. The Eq.~(\ref{eq_ghdenflu}) is then best fitted to this curve using non-linear least square fits resulting in optimal parameters for each mode. A fitted curve for $k_4$ mode is also shown in Fig.~\ref{fig:ckt_fit} (dashed lines) which closely follows the DAF from MD. 

\subsection{Comparison with MD and discussion}
In the previous subsection, the DAF generated through MD data is found to fit well with the Eq.~(\ref{eq_ghdenflu}). This fitting has been performed by fixing the  various transport parameters such as $c_s$, $\Gamma_s$, $\gamma$ and thermal diffusivity $a = \lambda/\rho c_p$ as fitting parameters. As the fitting procedure involves multiple parameters, it should be noted that not all of them can be varied arbitrarily.  Firstly, the parameter $c_s$ explicitly depends only on the frequency of the DAF time series and hence gets  decoupled  from the others. The quantity $A$ independently appears in the exponential of the first term and it also appears in the exponential of the second term (corresponds to the viscosity constant). So these two terms ($A$ and the viscosity term) cannot be arbitrarily chosen to fit the MD data. 

\par
\begin{figure}[h]
\includegraphics[width=\linewidth]{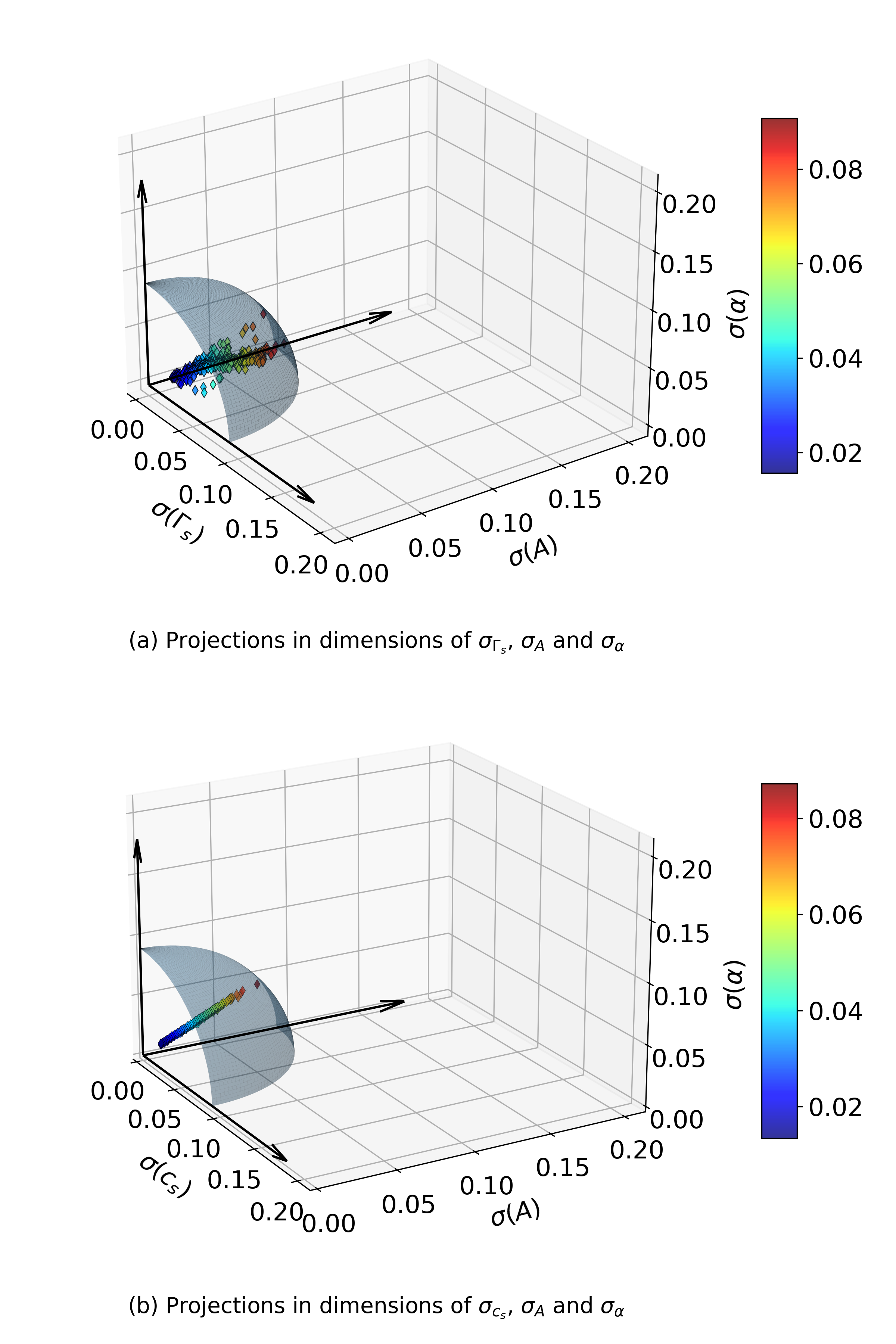}
\caption{3D projections of MSDs from the parameter hyperspace in normalized units. The colors of the points represents the Euclidean norms and shaded spherical surface encloses the region where $\sigma < \sigma_{max}$.}
\label{fig:ckt_error}
\end{figure}

 In order to have better confidence in the fitting procedure, an exercise involving the statistical uncertainties is also performed. A multidimensional space for the statistical errors around the fitted parameters is constructed. The statistical errors are quantified in terms of Mean Squared Deviations (MSD) of each parameter. As we have four fitting parameters for each $k$, the number of dimensions of this space will be four. These values are independently calculated for each $k$ and the maximum MSDs for individual parameters are collected. The spread in the unit of fractions (ratio of the error to value of the parameter) are denoted as  $ \pm \sigma$ with $\sigma_{max} = [0.0003,  0.049, 0.077, 0.041]$ corresponding to $[c_s, A, \Gamma_s, \alpha ]$. Here $\alpha$ is the coefficient of the first term in the Eq.~(\ref{eq_ghdenflu}) as defined for Eq.~(\ref{eq:gkformula}). To visualize the extent of deviations,  the MSDs are shown in 3D projections of the four-dimensional hyperspace of parameters  in Fig.~\ref{fig:ckt_error} for all modes. The Fig.~\ref{fig:ckt_error}(a) shows the populations of MSDs estimated in the parametric space of $\sigma_{\Gamma_s}$, $\sigma_A$ and $\sigma_{\alpha}$ dimensions. Similar information is shown in Fig.~\ref{fig:ckt_error}(b) corresponding to $\sigma_{c_s}$, $\sigma_A$ and $\sigma_{\alpha}$ dimensions. The colors of each scatter point show the Euclidean norm of the point which conveys the maximum possible deviation of all dimensions combined. It is evident from the figures that the populations of MSDs are limited to a small region within the parametric hyperspace, hence the fitting procedure is statistically accurate. 

\par In addition to ensuring the statistical accuracy of the fitting procedure we have also carried out an independent check on the validity of the estimated transport coefficients by comparing them with values available through various models in the literature. The comparison presented below covers  Yukawa systems in 2D and 3D along with a discussion on the physical  effects of strong coupling terms and the memory effects on the transport parameters. 

\par Before going to a one-by-one comparison, it is important to check the reduction of Eq.~(\ref{eq_ghdenflu}) in some important asymptotic limits. 
For an ideal uncharged fluid $\omega_p =0$ or $\kappa \rightarrow \infty$ Eq.(\ref{eq_ghdenflu}) exactly reduces to the DAF of an ideal fluid \cite{Hansen2013} as shown below

\begin{eqnarray}
\left \langle \rho_k (t) \rho_{-k}(0) \right\rangle =&& \left( \frac{\gamma-1}{\gamma} \right)\exp(-D_Tk^2t)+\nonumber \\
&&\frac{1}{\gamma}\exp(-\Gamma_sk^2t)\cos(c_skt)
\label{eq_denflu}
\end{eqnarray}
where 
$$
\Gamma_s = \frac{1}{2}\left(b+a\frac{\gamma-1}{\gamma}\right) 
$$
\par A comparison between Eqs.~(\ref{eq_ghdenflu}) and (\ref{eq_denflu}) shows that the transport terms such as $c_s$, $\gamma$ and thermal conductivity are modified through a new form of compressibility. While the longitudinal viscosity appearing in $\Gamma_s$ is modified through a term containing the relaxation time $\tau_m$.

The speed of acoustic modes \cite{Rao1990a} in Yukawa systems can be estimated using various methods and models including molecular dynamic (MD) simulations~\cite{Ohta2000}, QLCA \cite{Kalman2005} and fluid models supplemented with an equation of state, etc. For estimations of the adiabatic constant, parametric equation of state obtained from MD simulations or other models, are used in some reported cases \cite{Khrapak2015,Khrapak2014}.  Among all these methods the QLCA approach requires high coupling regimes for charges to be localized \cite{Golden2000} and the fluid approach is reported to be accurate in $\kappa \leq 3$ regimes \cite{Semenov2015}. Also, a direct experimental implementation is difficult for all the above cases. 

The following important point related to the expression for sound speed using the QLCA method by \citet{Kalman2000} is worth noting here. According to Eq.~(19) of reference [\onlinecite{Kalman2000}] the approximate expression of longitudinal phase velocities of Yukawa Systems for the limit of $k \rightarrow 0$ is given as 

\begin{equation}
s_L^2 = \omega_p^2a_{ws}\left[f(\kappa) + \frac{1}{\kappa^2}\right]
\label{eq:qlca_cs}
\end{equation} 
with $f(\kappa)$ as a fitting function. 
The expression obtained through present derivation as in Eq.~(\ref{eq:sound_speed}) is also in a similar form but with an explicit $k$ dependence and a more physically meaningful $k$ independent term, compressibility $K_T$. Thus the present form avoids the need for ambiguous parametric fitting on the estimation of sound speed. A similar form of dispersion relation is also reported in other places, for example, see Ref~\citep{Kaw1998} and references therein.

\begin{figure}
\includegraphics[width=\linewidth]{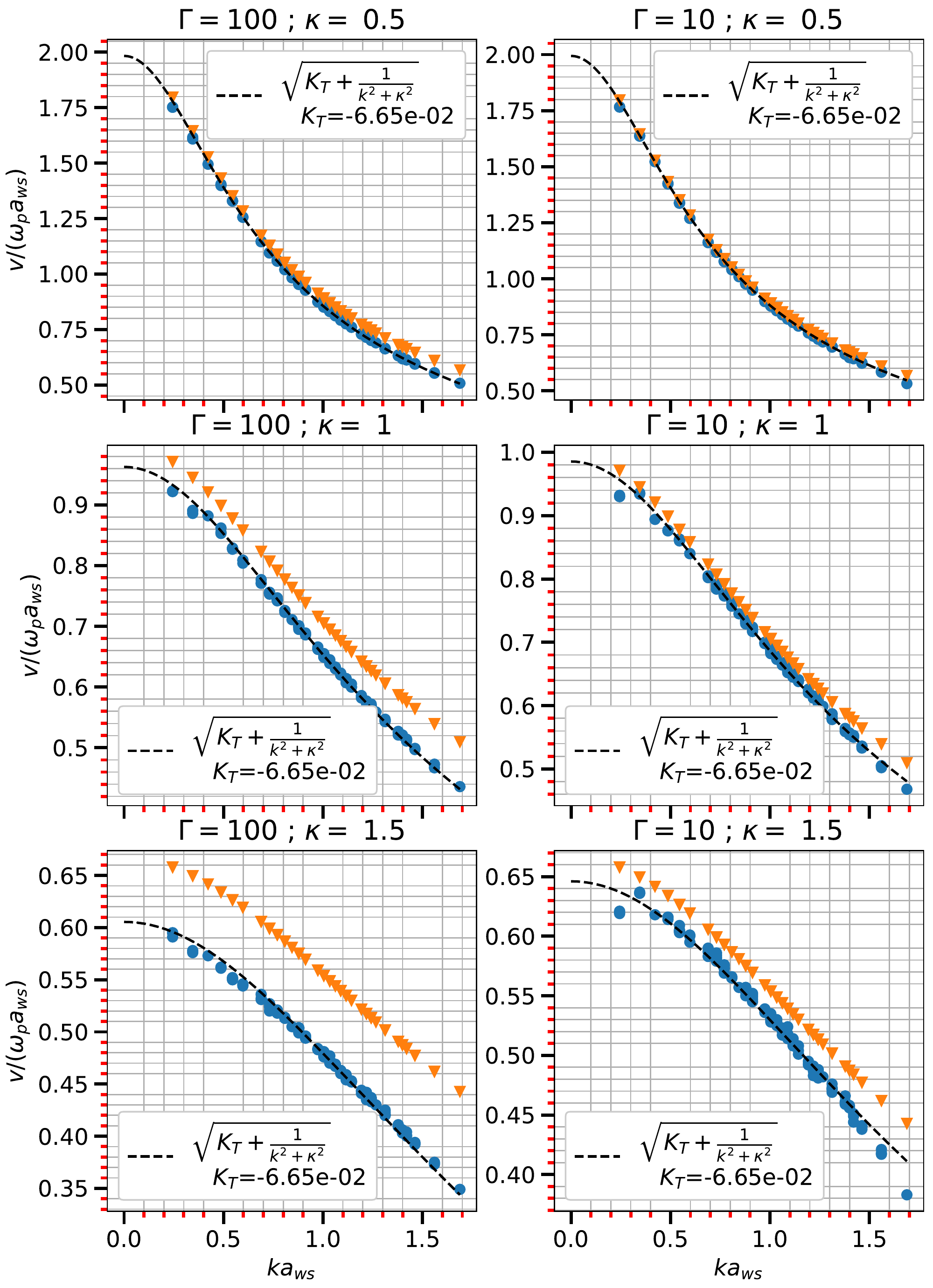}
\caption{Sound Speed ($c_s$) with $ka_{ws}$ for different $\Gamma$ and $\kappa$ in a 3D system. The inverted triangles represent $\sqrt{1/(k+\kappa^2)}$} term and dashed line shows the fitting with analytic expression for sound speed from Eq. (\ref{eq:sound_speed}).
\label{cs_fitting-3d} 
\end{figure}

\begin{figure}
\includegraphics[width=\linewidth]{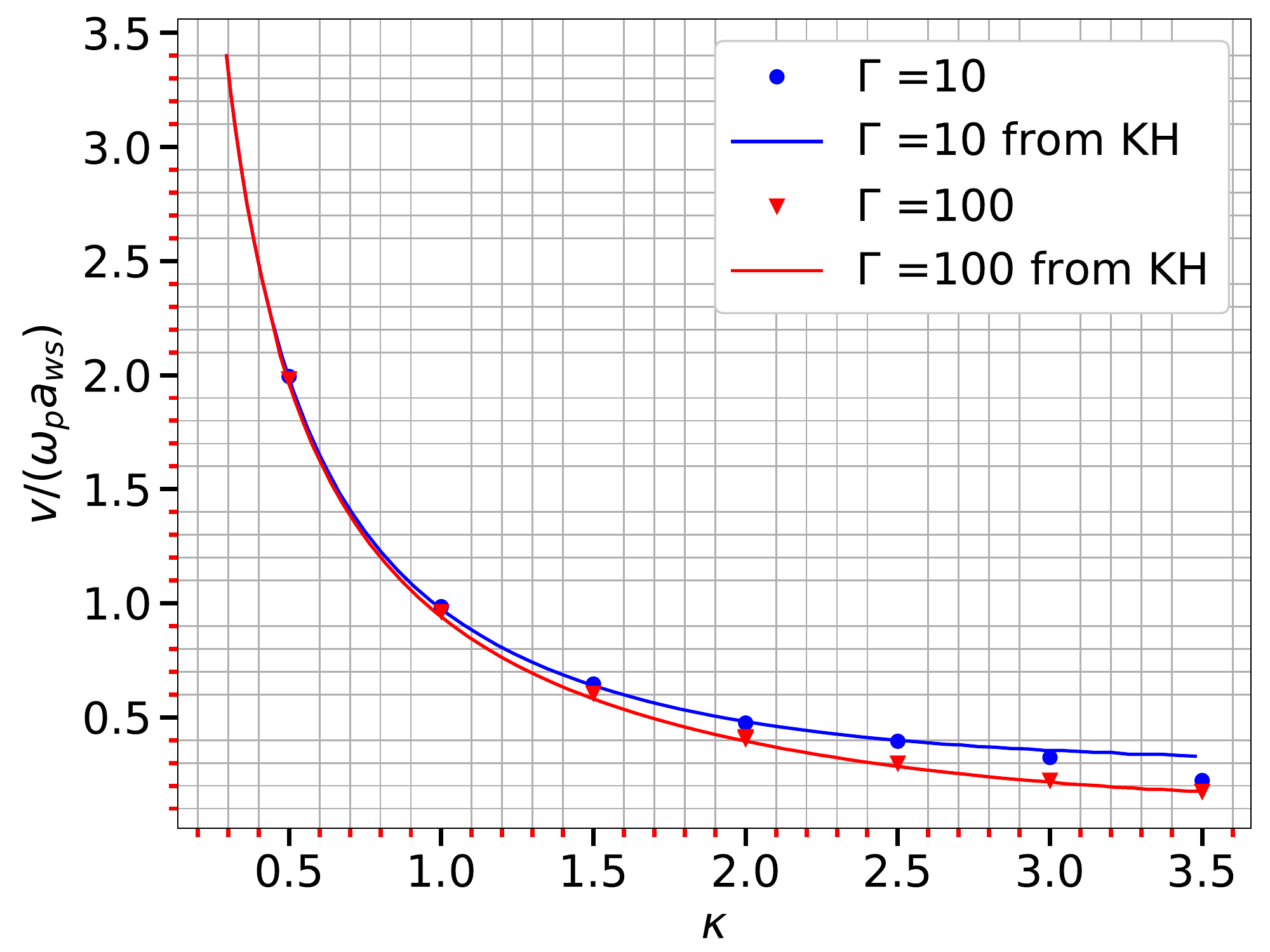}
\caption{Comparison of sound speed obtained in $k\rightarrow 0$ limit for a 3D system with results from \citet{Khrapak2015}. }
\label{cs_compare_3d}
\end{figure}

Now the left hand side of Eq.~(\ref{eq:sound_speed}) can be obtained from fitting Eq.~(\ref{eq_ghdenflu}) with MD data for each wave-vector $k$. These values can be further fitted with the expression in right hand side of Eq.~(\ref{eq:sound_speed}) as shown in Fig~\ref{cs_fitting-3d}. The fitting procedure is also capable of separating the wavelength dependent term from the other term in expression. The circles in the Fig.~\ref{cs_fitting-3d} show MD point for the left hand side of Eq.~(\ref{eq:sound_speed}) and broken lines show the fit using the expression in the right hand side. The term $\sqrt{1/(k+\kappa^2)}$ is shown with inverted triangle and calculated values of $K_T$ is also 
mapped in Fig.~\ref{cs_fitting-3d}. The values for $c_s$ are then extrapolated to $k\rightarrow 0$ and compared with results of ~\citet{Khrapak2015} in Fig.~\ref{cs_compare_3d}. The figure shows the comparison for two values of $\Gamma$s and different values of $\kappa$ ranging from 0.5 to 3.5. It should be noted that values by \citet{Khrapak2015} are accurate up to $\kappa = 3$ and the comparison is performed for a 3D dusty plasma.


\begin{figure}
\includegraphics[width=\linewidth]{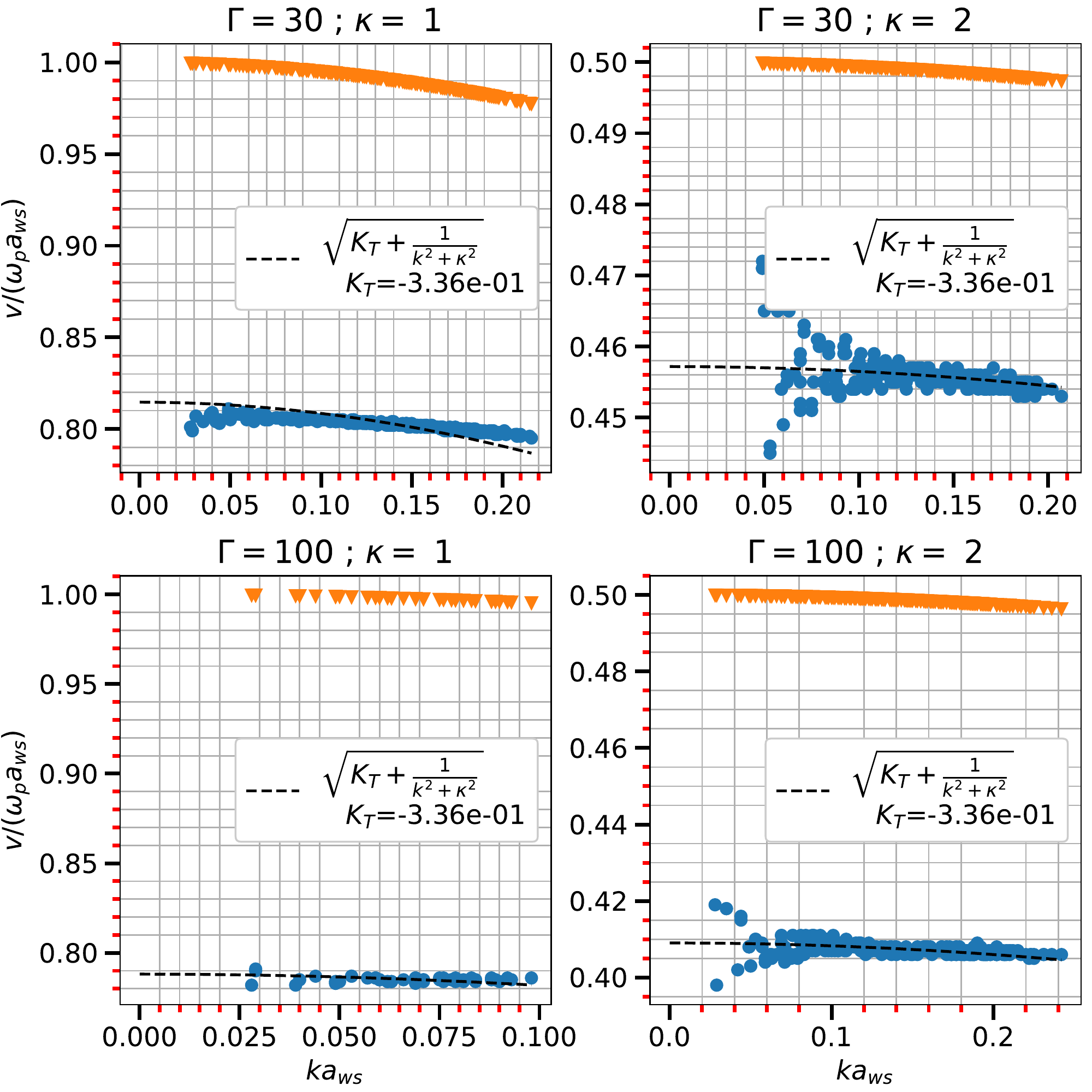}
\caption{Sound Speed ($c_s$) with $ka_{ws}$ for different $\Gamma$ and $\kappa$ in a 2D system. The inverted triangles represent $\sqrt{1/(k+\kappa^2)}$  and dashed line shows the fitting with analytic expression for sound speed from Eq.~(\ref{eq:sound_speed})}
\label{cs_fitting_2d}
\end{figure}

\begin{figure}
\includegraphics[width=\linewidth]{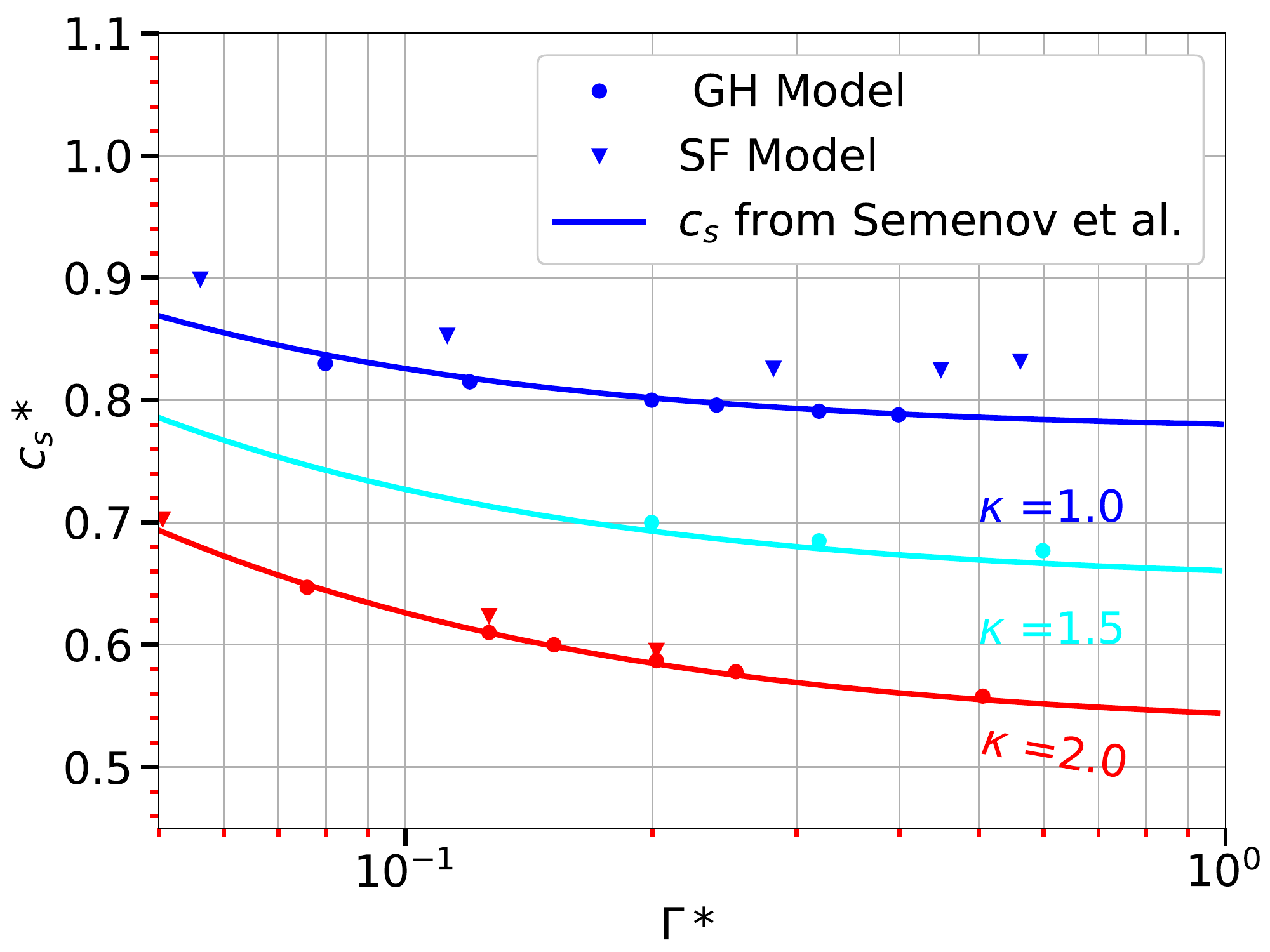}%
\caption{Comparison of sound speed as a function of coulomb coupling parameter from present work (triangles) with \citet{Semenov2015} (solid lines). Following the same normalization used in Ref~[\onlinecite{Semenov2015}], the sound speed is normalized as $c_s* = c_s\kappa/(\omega_{pd}a_{ws})$ and  effective coulomb coupling parameter as $\Gamma *=\Gamma/\Gamma_m$ where $\Gamma_m(\kappa)=131/[1-0.388\kappa^2+0.138\kappa^3-0.0138\kappa^4]$.}
\label{fig:cs}
\end{figure}
\begin{figure}
\includegraphics[width=\linewidth]{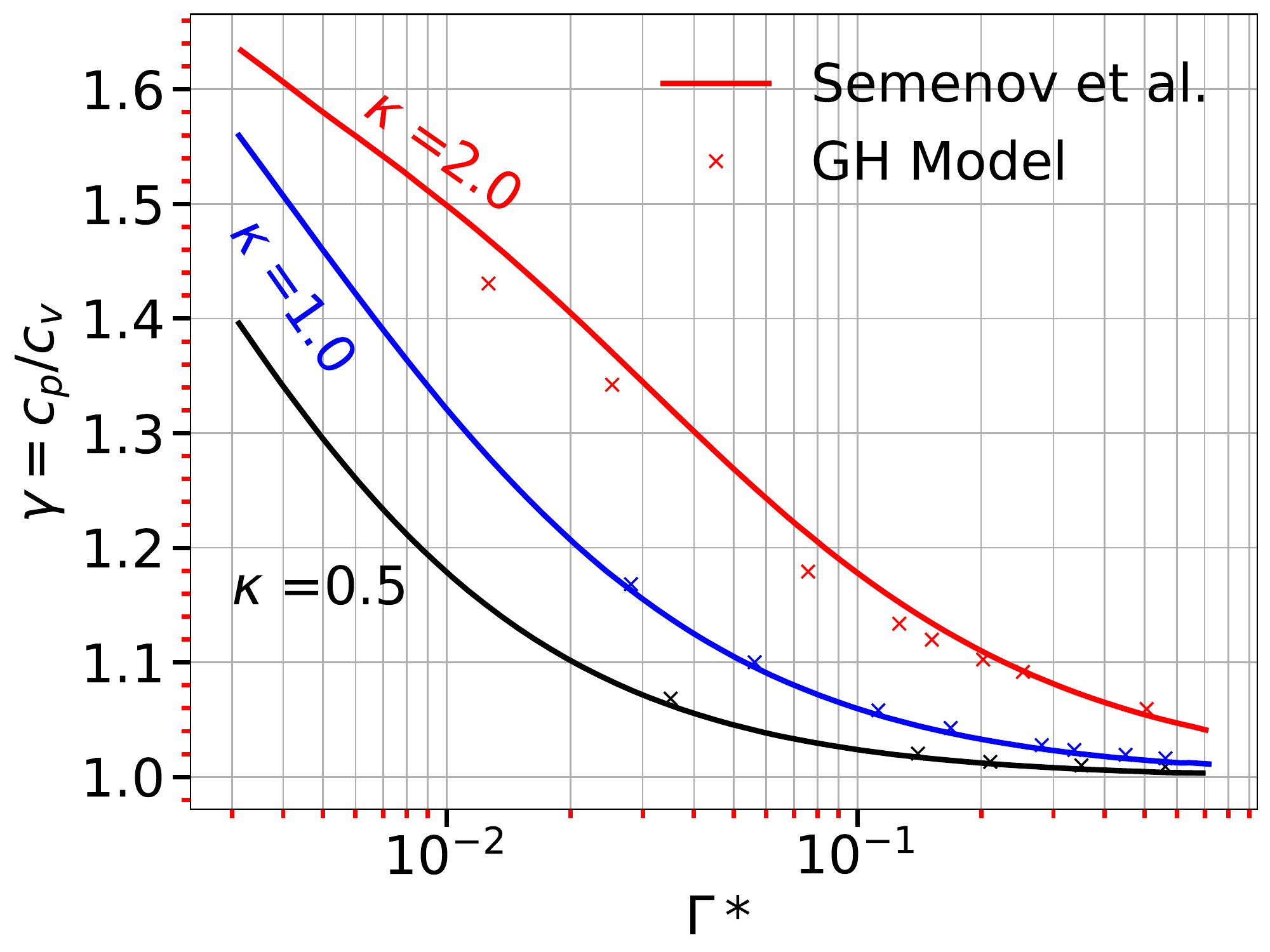}%
\caption{Comparison of adiabatic constant ($\gamma$) as a function of coulomb coupling parameter from present work (crosses) with \citet{Semenov2015} (solid lines). Same normalization for Coulomb coupling parameter as in Fig.~\ref{fig:cs} is used here also.}
\label{fig:gamma}
\end{figure}

To check the validity of the present model for 2D dusty plasma, the following comparisons are performed. A plot for 2D cases exactly similar to Fig.~\ref{cs_fitting-3d} is shown in Fig~\ref{cs_fitting_2d}. Similarly, the sound speed estimated for 2D cases and a comparison of present calculations
with \citet{Semenov2015} is shown in Fig~\ref{fig:cs}. The solid lines are from Ref [\onlinecite{Semenov2015}] and circles are from the present calculations using a combination of MD data, Eqs.~(\ref{eq_ghdenflu}) and (\ref{eq:sound_speed}). The inverted triangles are for the limiting case of simple fluids as in Eq.~(\ref{eq_denflu}). Both the axes are normalized as described in the caption to make it analogous to the work of  ~\citet{Semenov2015}. The comparisons are presented for three cases with  $\kappa$=0.5, 1, 2 and for many values of 
$\Gamma$  from 1 to 100.
Following important points can be noted from Fig.~\ref{fig:cs}. Firstly, the present model agrees well with the results of \citet{Semenov2015}. Secondly, as $\kappa$ increases the values calculated with the GH model (Eq.~\ref{eq_ghdenflu})  approach the values estimated using the simple fluid model. As discussed earlier this point is also  in line with expectations.
This in turn validates the present derivation.
A similar comparison for adiabatic constant $\gamma$ is also given in Fig.~\ref{fig:gamma}. It should be noted that for calculating $\gamma$, an equivalence  between the quantity $(\gamma - 1)/\gamma K_T$ for the cases  with  and without background, as explained by ~\citet{Salin2007}, is used. Here also the present method can estimate values closer to that of \citet{Semenov2015} even for $\gamma$s close to one.

From the above discussions, it is clear that the sound speed and gammas obtained using the present model and MD data in rigorous ways agree with other available models in the literature even though they are very different from the present approach. For completeness, in the rest of this section a comparison of another important transport parameter, the thermal conductivity is presented.

The thermal conductivity estimation can be done by equilibrium MD simulations using Green-Kubo formula, which is based upon the fluctuation-dissipation theorem \cite{Kubo1957}. This method involves the computation of heat current auto-correlation which has a slow convergence \cite{Schelling2002}. The definition of local heat current is not unique and reported to cause statistical errors \citep{Marcolongo2020,Schelling2002}, and in many cases (for example, if the potential is not pairwise additive)  accurate estimation of thermal conductivity is not possible with GK method \cite{Schelling2002,Cheng2020}. Another method to calculate thermal conductivity is to use non-equilibrium molecular dynamics (NEMD) simulations \cite{Muller1997} by inducing a local temperature gradient in a small region of the system to estimate the heat flux ~\cite{Schelling2002}. In principle NEMD methods closely mimic the experimental situations and are not difficult to implement in MD, but there exist many computational issues~\cite{Schelling2002}. The limitations include finite-size effects and non-linear responses due to temperature gradient \cite{Bedrov2000}. A review of both equilibrium methods and non-equilibrium methods with merits and demerits are available in a study by ~\citet{Schelling2002}. However, both of the aforementioned methods are difficult to deploy in experimental Yukawa systems. For GK methods, experimentally estimating the local heat current is challenging, while in non-equilibrium methods creating a local thermal gradient and keeping the overall system at a constant average temperature, and measuring local heat flux is difficult.

\begin{figure}
\includegraphics[width=\linewidth]{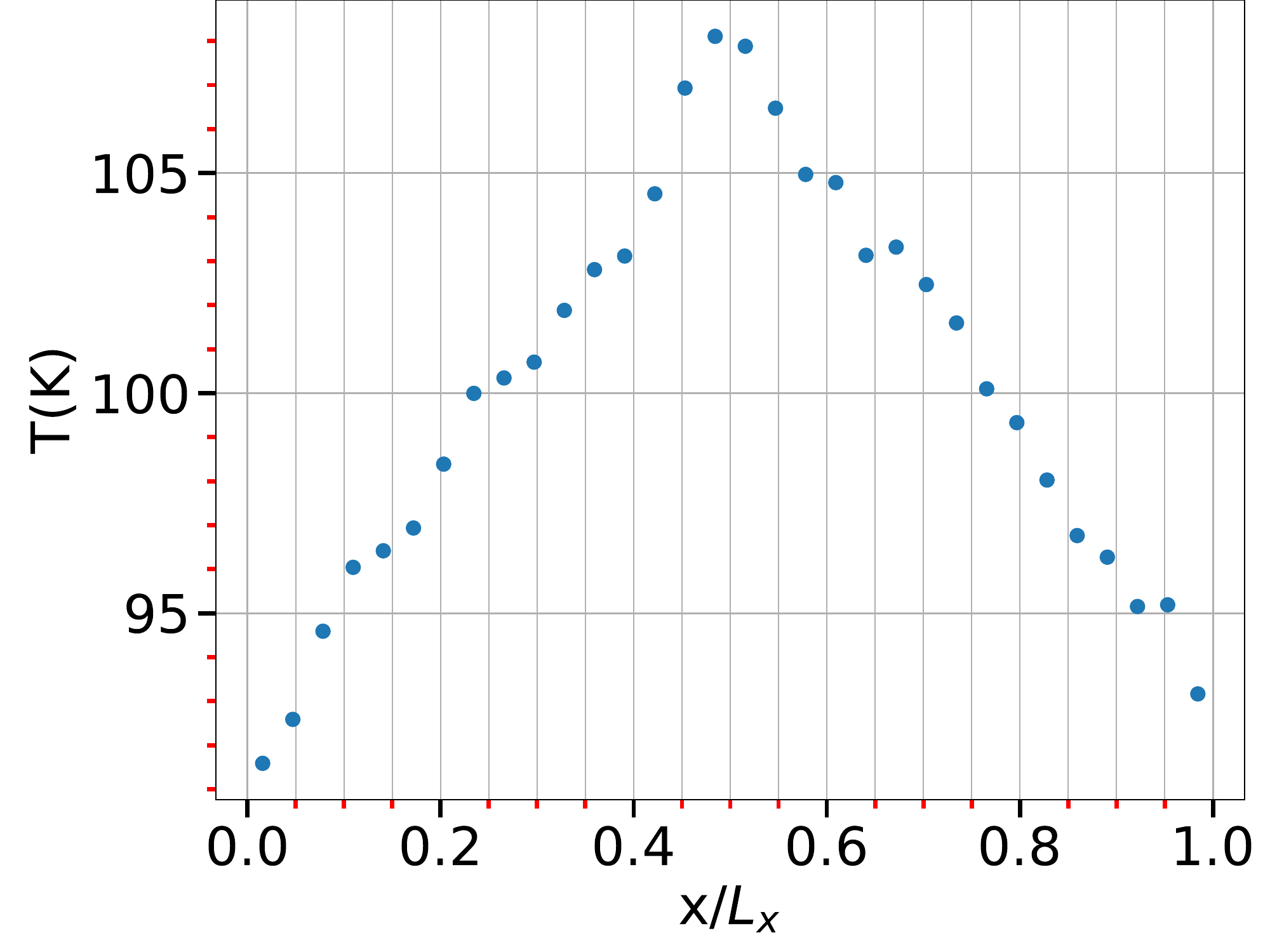}%
\caption{Temperature profile constructed for estimation of thermal conductivity from NEMD method. }
\label{fig:nemd}
\end{figure}

To validate the prediction of thermal conductivity using the present model, MD calculations for NEMD are separately performed as described below. For NEMD calculation, a reversible non-equilibrium method proposed by \citet{Muller1997}(MP) is used. It is based on the idea of deliberately imposing a heat flux and measuring the system response as a temperature gradient profile. The system is divided into 32 slabs along  $\hat{x}$ direction and heat flux is imposed by exchanging the kinetic energy of the \lq \lq coldest" particle in one slab with \lq \lq hottest" in another slab.  The induced temperature gradient,  as the response of the system, is measured by taking ensemble averages. The temperature profile after establishing the temperature gradient is shown in Fig.~\ref{fig:nemd}. Now, for a 2D system the thermal conductivity is related to heat flux using Fourier's law as
\begin{equation}
\lambda = \frac{E}{2Lt \left \langle \delta T/\delta x\right \rangle }
\label{eq:Fourier's law}
\end{equation}
where E is the total energy exchanged in time $t$, $L$ is length of slab and $\left\langle  \delta T/\delta x \right\rangle$ is the ensemble average of temperature gradient. 

\begin{figure}
\includegraphics[width=\linewidth]{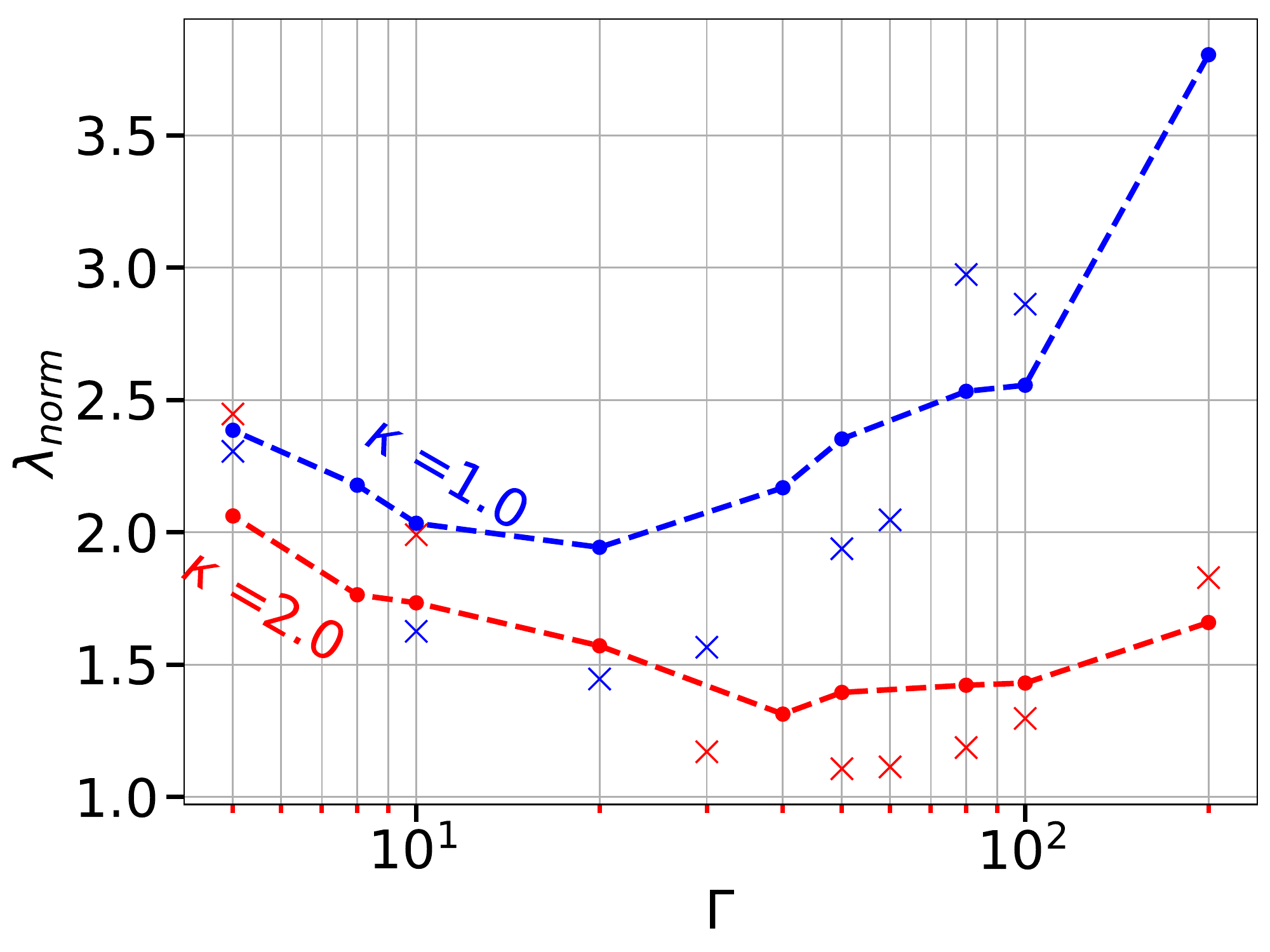}%
\caption{Comparison of heat conductivity calculated with NEMD Method (dashed lines),  and the present work using Eq.~(\ref{eq_ghdenflu}). The thermal conductivity is normalized with $\omega_p$ as $\lambda_{norm} = {\lambda}/{nk_{\beta}\omega_pa_{ws}^2}$}
\label{fig:lambda_all}
\end{figure}

The Eq.~(\ref{eq:Fourier's law}) is used to estimate the thermal conductivity from a known heat flux and temperature profile. A comparison of the results obtained using Eq.~(\ref{eq_ghdenflu}) with that obtained using NEMD for $\kappa=1$ and $\kappa=2$ for many values of $\Gamma$  is shown in Fig.~\ref{fig:lambda_all}. The present simulation agrees well with NEMD method considering the reported inaccuracy of NEMD method up to 20 $\%$.  

 As discussed earlier, the NEMD method has its own computational disadvantages. As the present method closely follows the analytical treatment and the DAF is calculated from particle fluctuations,  it is free from such problems but can be prone to statistical errors that arise from fitting procedures.

Validation of the final parameter in Eq.~(\ref{eq_ghdenflu}) namely $\Gamma_s$ is not performed here as the same expressed in the form of Eq.~(\ref{eq:Gamma_s}) is not available in literature.

 \par An important extension of this work could be an analytical estimation of the stress autocorrelation function (SACF) which is related to the viscosity using a Green-Kubo formula~\cite{Feng2011}. In this regard, the matrix equation (Eq.~\ref{eq:ghhydromatrix}) can be extended by incorporating the transverse currents ($j_x$,$j_y$). This would result in a hydrodynamic matrix of order 5, which can be inverted to approximately solve the system of equations resulting in analytical expressions of current densities. These current densities can be used along with the conservation law of momentum to calculate the stress tensor. 
\par One could also have an alternate approach to obtain viscosity using the auto correlation of the time derivative of the current density. For example, this can be achieved using the following form of expression, which has been previously used for the case of simple fluids.\cite{Hansen2013}

\begin{equation}
\eta = \frac{\beta m^2}{V}\lim_{k\rightarrow 0}\lim_{\omega\rightarrow 0} Re \int_0^\infty \frac{1}{k^2} \langle \bm{\dot{j}}^x_k(t) \bm{\dot{j}}^x_{-k}(0)\rangle \exp(\iota \omega t)dt. 
\end{equation}
 
\par  Furthermore, we would like to make the following important point. As shown above, the present approach of using Eq.~(\ref{eq_ghdenflu}) and MD simulations can be used for accurate estimation of various transport parameters in a single framework. Moving forward, as explained below, there exists an interesting possibility to replace MD data with experimental data. One of the beauties of laboratory dusty plasma systems is their simplicity in obtaining the particle trajectories using fast cameras~\cite{Hariprasad2018,Hariprasad2020,Jaiswal2016}. These particle trajectories can be used to obtain a DAF. The experimentally obtained DAF can then be matched with Eq.~(\ref{eq_denflu}) as discussed earlier. In other words, the particle trajectories obtained through MD simulation in the present work can be replaced with
experimental measurements. As discussed earlier, the experimental implementation of previous individual models for each thermodynamic quantities such as $c_s$,  $\gamma$, $\lambda $  are difficult, need more complicated diagnostics and more importantly, require separate treatments.  Using Eq.~(\ref{eq_ghdenflu}) of the present work with experimentally measured particle trajectories enables the estimation of many important transport parameters in a single framework. For normal systems like simple fluids, this cross-validation is not possible as experimentally measuring the individual particle dynamics and fluctuations is impossible. In short, the present work opens up a window to cross-validate the dynamics of microscopic fluctuations at  hydrodynamic limits with theoretical, computational, and experimental means. An experimental attempt for the same is presently under way and will be reported later.

\par Finally, we discuss an important issue related to the use of the Green-Kubo relation for the determination of a transport coefficient directly from experimental data. 

\begin{figure}
\includegraphics[width=\linewidth]{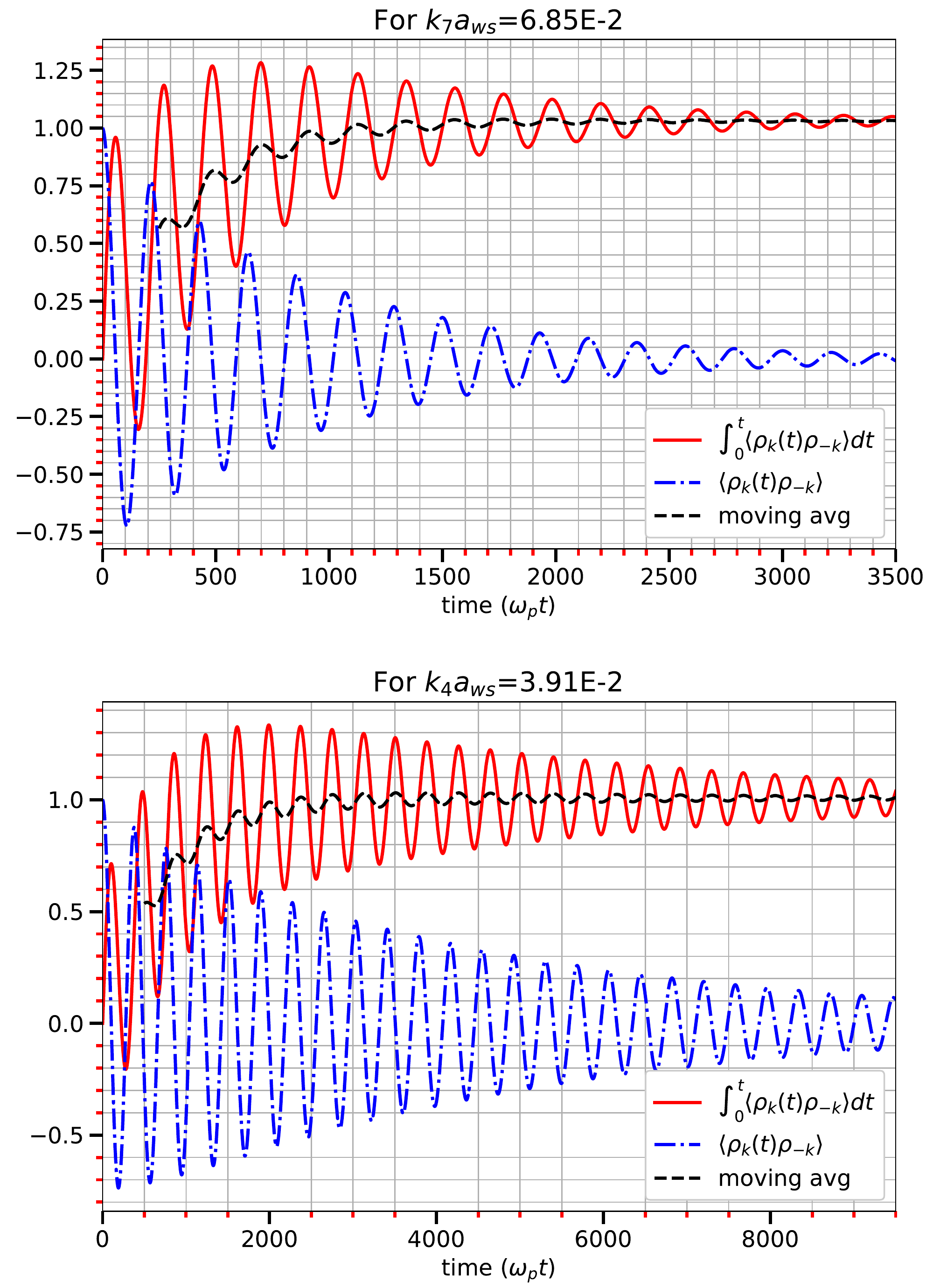}
\caption{The DAF and GK integral (Eq.~(\ref{eq:GKint})) with finite upper limit along with moving average for $k_4$ and $k_7$ mode from ~Fig.\ref{fig:ckt_fit}. The integration (solid red curve) is normalized to the asymptotic value}.
\label{fig:daf_int}
\end{figure}

The GK formula can be written as 
\begin{equation}
I = \int_0^\infty\left \langle \rho_k (t) \rho_{-k}(0)\right\rangle dt
\label{eq:gkformula}
\end{equation} 
where for the integrand we have used the density auto-correlation function. In an experiment the DAF is obtained as a finite time series and one needs to choose an upper limit of integration to evaluate the integral (\ref{eq:gkformula}).  Typically, the time corresponding to the first zero crossing of the time series has been chosen, e.g. in estimating the viscosity of a 2D dusty plasma using the stress autocorrelation experimental data \cite{Feng2013}. Here we discuss the appropriateness of such a choice for the DAF data. Since we have an analytic expression for the DAF the GK integral can be evaluated analytically for arbitrary values of the upper limit. The analytic expression is given by,
\begin{eqnarray}
I = && \int_0^t\left \langle \rho_k (t) \rho_{-k}(0)\right\rangle dt=\frac{\alpha}{Ak^2}\left[1-\exp(-Ak^2t)\right] +\nonumber \\ 
&&\frac{(1-\alpha)\Gamma_s}{A^2k^2+c_s^2}\left[ 1+\exp(-\Gamma_sk^2t)\left(\frac{c_s\sin(c_skt)}{\Gamma_sk^2}-\cos(c_skt) \right)\right] \nonumber \\
\label{eq:GKint}
\end{eqnarray}
where 
$$\alpha =  \left(\frac{K_T(\gamma-1)/\gamma}{K_T+\frac{\omega_p^2}{k^2+\lambda_D^{-2}}} \right)$$
For $t\rightarrow \infty$,  Eq.~(\ref{eq:GKint}) gives
\begin{equation}
I = \frac{\alpha}{Ak^2} + \frac{(1-\alpha)\Gamma_s}{A^2k^2+c_s^2}
\label{GK_asymp}
\end{equation}
which is the exact value of the GK formula. The question is, for what choice of the upper limit $t$ does one get a value of the integral (\ref{eq:GKint}) that is reasonably close to RHS of Eq.~(\ref{GK_asymp}). To answer this question we have numerically plotted Eq.~(\ref{eq:GKint}) as a function of $t$ for two different  DAF curves. These are shown in Fig.~\ref{fig:daf_int} where the corresponding DAF curves have also been plotted. 

As can be seen, the choice of the first zero crossing of the time series as the upper limit $t$, can give a grossly inaccurate value of the integral (\ref{eq:GKint}).  One needs to go much further depending upon how fast the envelope of the DAF is decaying to get a value that is reasonably close to the actual value.\\

\section{Summary}

In the present work, an analytical relation for the time dynamics of DAF for a Yukawa fluid has been explicitly derived. This analytical form is then used directly for the estimation of various transport coefficients using GK relations. This analytical form can also be compared directly with experimental or MD data to obtain important transport coefficients using proper fitting procedures. A potential generalization of the present work is to extend the calculations by including transverse current density components to obtain useful analytical expressions for other important parameters like the stress-autocorrelation function. Furthermore, the expression of the DAF as in Eq.~(\ref{eq_ghdenflu}) can be usefully employed to estimate an upper limit of integration of the various GK formulae, when the DAF is obtained as finite time series from  experimental data.  The important limitations of the present work are given below.
\begin{enumerate}
\item The present model and Eq.~(\ref{eq_ghdenflu}) are valid only for Yukawa systems in the hydrodynamic limit.
\item Electromagnetic effects have been neglected in the present work.
\item The validity of the present calculations are limited to longer wavelengths as we have used O($k^2$) approximation. In other words, the wavelengths in consideration should be larger than the mean free paths for the applicability of the hydrodynamic regime. Mathematically $kl_c <<  1$ where $l_c$ is the scale length of collisions.  
\item Neutral drags and other damping terms have also been neglected.

\end{enumerate}

\begin{acknowledgments}
A.S. is grateful to the Indian National Science Academy (INSA) for the INSA Honorary Scientist position. The MD simulations described in this paper were performed on ANTYA, an IPR Linux Cluster.
\end{acknowledgments}

\appendix
\section{\label{sec:appendix} Method of solving equation~(\ref{eq:detHLeq})}
The roots of a quartic equation such as Eq.~(\ref{eq:detHL}) of the form
\begin{equation}
P(k)z^4 +Q(k)z^3 +R(k)z^2 +S(k)z^1+T(k) = 0
\label{eq:polynomial}
\end{equation}
with the coefficients P, Q, R, S and T being functions of $k$, can be approximately estimated using a power series method. In this method, a trial solution of the form
$z_t = z_{t0} +z_{t1}k+z_{t2}k^2...$ is substituted in equation. Then the terms with same order of $k$ are collected together and the coefficient $z_{t0}$ is estimated by considering the lowest order terms in $k$. For Eq.~(\ref{eq:detHL}), the lowest order coefficient of trial solution is
\begin{equation}
z_{t0}= 0,0,0,-1/\tau_m
\label{eq:zt0}
\end{equation}
Now, this process is repeated again with substitution of trial solution with the calculated $z_{t0}$ from previous step to estimate next order coefficient i.e. $z_1$.
For each value of $z_{t0}$, the order (k) coefficient of trail solution is
\begin{equation}
z_{t1} = 0,\pm \sqrt{K_T+\frac{\omega_p}{k^2+\lambda_D^{-2}}},0
\label{eq:zt1}
\end{equation}
This is repeated until required approximation in order of $k$, which is $O(k^2)$ in this paper, is reached. The second order coefficient of the trial solution is
\begin{eqnarray}
&&z_{t2} =-a \left ( 1 + \frac{K_T(1/\gamma -1)}{K_T + \frac{\omega_p^2}{k^2+\lambda_D^{-2}}}\right ),\nonumber \\
&&-\frac{1}{2}\left(b+a\frac{K_T(1-1/\gamma)}{K_T + \frac{\omega_p^2}{k^2+\lambda_D^{-2}}} -\frac{\omega_p^2\tau_m}{k^2+\lambda_D^{-2}} \right) ,\left[b-\frac{\omega_p^2 \tau_m}{k^2+\lambda_D^{-2}}\right] \nonumber \\
\label{eq:zt2}
\end{eqnarray}

The final approximate roots of order $O(k^2)$ are then estimated from Eqs. (\ref{eq:zt0}-\ref{eq:zt2}) as shown in section \ref{sec:model}.

\end{document}